\documentclass[a4paper]{jpconf}
\usepackage{iopams}
\usepackage{graphicx}
\usepackage{xcolor}

\newcommand{\ddt}[0]{\frac{\partial}{\partial t}}
\newcommand{\ddtau}[0]{\frac{\partial}{\partial \tau}}

\newcommand{\mbf}[1]{\mathbf{#1}}
\renewcommand{\k}[0]{\mbf{k}}

\begin{document}
\title{Non-Markovian Effects in the Spin Transfer Dynamics in 
Diluted Magnetic Semiconductors due to Excitation in Proximity to the
Band Edge}
\author{M.~Cygorek, V.~M.~Axt}
\address{ Theoretische Physik III, Universit{\"a}t Bayreuth, 95440 Bayreuth, Germany}
\ead{Moritz.Cygorek@uni-bayreuth.de}

\begin{abstract}
The non-Markovian effects in the spin dynamics in diluted magnetic 
semiconductors found in quantum kinetic calculations can be reproduced
very well by a much simpler effective single electron theory, if a
finite memory is accounted for. The resulting integro-differential equation
can be solved by a differential transform method, yielding the Taylor series
of the solution. From the comparison of both theories it can be concluded that
the non-Markovian effects are due to the spectral 
proximity of the excited electrons
to the band edge.
\end{abstract}

\section{Introduction}
Diluted magnetic semiconductors (DMS) are a class of workhorse materials in the
field of semiconductor spintronics, since they combine the magnetic degree
of freedom with the versatility and highly developed fabrication schemes
of the semiconductor technology.
Usually, Mn doped II-VI or III-V semiconductors are studied and a
localized $s$-$d$ interaction between the carrier and Mn spins
modelled by a Kondo-like Hamiltonian has been found to describe the
magnetic properties and the spin dynamics of DMS very well.

A numerical calculation based on a quantum kinetic theory (QKT) for the 
spin dynamics in DMS governed by the $s$-$d$ interaction\cite{Thurn12} 
showed that, among other phenomena,
non-Markovian effects, such as overshoots or oscillations of the total 
spin polarization, can be found in one- and two-dimensional 
systems\cite{Thurn13_1,Thurn13_2}.
The quantum kinetic theory can be presented
in a more easy-to-use and intuitive way, by eliminating the correlations at
the cost of a memory integral. Because it was found that, in doing so, it
is crucial to account for a precession-like dynamics of the carrier-impurity
correlations, the equations are referred to as \textit{precession of
electron spins and correlations} (PESC) equations\cite{PESCarxiv}.
In the present article, we show that the non-Markovian spin dynamics in DMS,
found in the quantum kinetic theory, can be well described by an 
approximation of the PESC equations. The resulting integro-differential 
equation can be solved by a differential transform method (DTM)\cite{ZhouDTM}.
An analysis based on this simplified approach reveals
 that the non-Markovian effects are due to the proximity of the
electronic excitations to the band edge.

\section{Equation of motion}
In Ref.~\cite{PESCarxiv} effective equations of motion for the 
correlation-induced spin dynamics in DMS were derived. For initially vanishing 
magnetization of the magnetic impurities, the time evolution
of the conduction band electron spin polarization in a DMS quantum 
well structure can be found from Eq.~(7a) of Ref.~\cite{PESCarxiv}:
\begin{eqnarray}
\ddt s_{\omega_1}(t)=-\frac{\eta}{\pi}\int\limits_0^{t}dt'
\int\limits_0^{\omega_{BZ}}d\omega \cos[(\omega_1-\omega)(t-t')]
\bigg[s_{\omega_1}(t')+\frac 14 \big(s_\omega(t')-s_{\omega_1}(t')\big) \bigg],
\label{eq:anf}
\end{eqnarray}
where $s_{\omega_1}$ is the mean electron spin of electrons with 
energy $\hbar\omega_1$ (relative to the band minimum), 
$\eta$ is the spin transfer rate in the
Markov limit and 
$\hbar\omega_{BZ}$ is the energy at the end of the first Brillouin zone.
If we assume a parabolic band structure, we find 
$\omega_1=\frac{\hbar k_1^2}{2m^*}$ with effective mass $m^*$ for 
an electron with wave vector $\k_1$ and 
$\eta=\frac{35}{12}\frac{J_{sd}^2 m^*n_{Mn}}{\hbar^3 d}$ with coupling
constant $J_{sd}$, magnetic ion density $n_{Mn}$ and quantum well width $d$.

It is noteworthy that in the time derivative for the total spin, where
Eq.~(\ref{eq:anf}) is integrated over $\omega_1$, the term
$\big(s_\omega(t')-s_{\omega_1}(t')\big)$ cancels. 
Since this term can be expected to lead to an insignificant contribution
to the total spin, we henceforth neglect this term which simplifies the
analysis of the spin dynamics drastically.
Despite this argument being valid only for the total spin, we shall show by 
numerical calculation that also the individual spin dynamics for an
electron at the energy $\omega_1$ is reasonably well described by this 
approximation (cf. Fig 1(c) and (d)). 
Thus, Eq.~(\ref{eq:anf}) can be reduced to

\begin{eqnarray}
\ddt s_{\omega_1}(t)=-\frac{\eta}{\pi}\int\limits_0^{t}dt'\bigg[
\frac{\sin((\omega_{BZ}-\omega_1)(t'-t))}{t'-t}+
\frac{\sin(\omega_1(t'-t))}{t'-t}\bigg]
s_{\omega_1}(t')
\label{eq:idgl}
\end{eqnarray}

The phyiscal meaning of Eq.~(\ref{eq:idgl}) becomes most obvious when the
Markov limit is regarded, which assumes that $s_{\omega_1}$ changes on a
much slower timescale than the oscillations of the integral kernel.
Then, on the r.~h.~s. of Eq.~(\ref{eq:idgl}), 
$s_{\omega_1}(t')$ can be evaluated at $t'=t$ and drawn out of the integral. 
Keeping in mind that $\lim_{\omega\to\infty} \sin(\omega t)/t=\pi\delta(t)$
and that the integral ranges only over one half of the $\sin(\omega t)/t$ peak,
one finds:
$\ddt s_{\omega_1}=-\eta s_{\omega_1}$,
which shows a simple exponential decay of $s_{\omega_1}$ with the rate $\eta$.
This corresponds to a golden rule-type transfer of the electron spin 
to the impurity system.

However, the condition for the applicability of the Markov limit was
$\eta\ll \omega_1$ and $\eta\ll \omega_{BZ}-\omega_{1}$. For realistic parameters
(e.~g., the parameters used in Ref.~\cite{Thurn13_1,Thurn13_2} yield 
$\hbar\eta\approx 0.45$ meV) 
and excitations far away from the end of the first Brillouin zone,
only the latter condition is fulfilled, while for excitations 
close to the band edge,
$\omega_1$ can be of the same order of magnitude as $\eta$.
Thus, we apply the Markov limit ($\omega_{BZ}\to\infty$) 
only on the first term of Eq.~(\ref{eq:idgl}).
The number of parameters can be reduced by 
substituting $\tau:=\eta t$ and 
$\xi:=\omega_1 / \eta$. Then, the problem is transformed to:
\begin{eqnarray}
\ddtau \Phi_\xi(\tau)=-\frac12 \Phi_\xi(\tau)
-\frac{1}{\pi}\int\limits_0^{\tau}d\tau'
\frac{\sin(\xi(\tau'-\tau))}{\tau'-\tau}\Phi_\xi(\tau'),\quad \Phi_\xi(0):=1,
\label{eq:idgl3}
\end{eqnarray}
where $s_{\omega_1}(t)=s_{\omega_1}(0)\Phi_{\omega_1/\eta}(\eta t)$.
Thus, the shape of the time evolution depends only on the ratio 
between $\omega_1$ and $\eta$.
\section{Numerical Evaluation of the non-Markovian Spin Dynamics}
We solve the integro-differtial equation~(\ref{eq:idgl3})
by a technique similar to Zhou's 
differential transform method (DTM)\cite{ZhouDTM}, which consists of
Taylor-expanding all terms in Eq.~(\ref{eq:idgl3}) at $\tau=0$.
This yields a recursion relation between the derivatives of $\Phi_{\xi}$:
\begin{eqnarray}
&\Phi_\xi^{(i)}=-\frac 12 \Phi_\xi^{(i-1)}-
\frac 1\pi \sum\limits_{0\le 2m\le i-2} \frac{(-1)^m}{2m+1} \xi^{(2m+1)}
\Phi_\xi^{(i-2-2m)},
\label{eq:recrel}
\end{eqnarray}
where  $\Phi_\xi^{(i)}$ is the $i$-th derivative of $\Phi_\xi$ evaluated at
$\tau=0$.
The numerical evaluation of Eq.~(\ref{eq:recrel}) is very efficient and
$\Phi_\xi(\tau)$ can be calculated to high orders 
by substituting the
derivatives into the Taylor expansion. We refer to this algorithm as
the DTM calculation.

It is noteworthy that from the recursion relation~(\ref{eq:recrel})
closed expressions can be derived for $\Phi_\xi(\tau)$ to a certain 
order in the ratio $\xi$
by combinatoric analysis of the paths from $\Phi_\xi^{(0)}=1$ 
to $\Phi_\xi^{(n)}$ 
and comparing the Taylor series with that of known functions.
E.~g., to second order in $\xi$, we find:
\begin{eqnarray}
&\Phi_\xi(\tau)=e^{-\frac\tau 2}+
\frac\xi\pi\big[(2\tau+4)e^{-\frac\tau 2}-4\big]+
\bigg(\frac\xi\pi\bigg)^2\bigg[(2\tau^2+16\tau+48)e^{-\frac\tau 2}+8\tau
-48\bigg]+\mathcal{O}(\xi^3)
\label{eq:snOrd}
\end{eqnarray}
which should be valid for excitations near the band edge where $\xi\ll1$ can
be fulfilled.
\section{Results}
\begin{figure}
\begin{minipage}[t]{\textwidth}   
\includegraphics{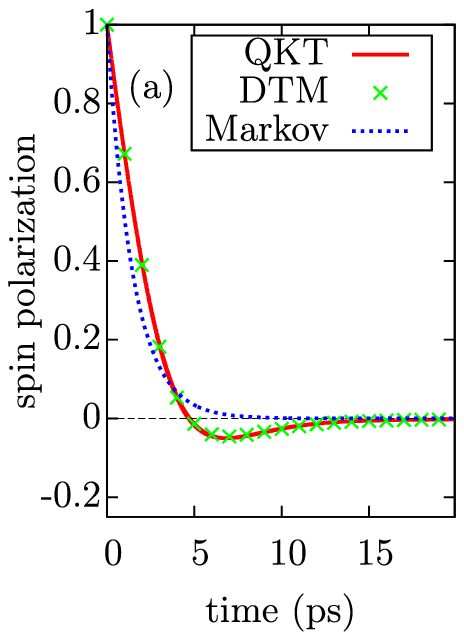}\includegraphics{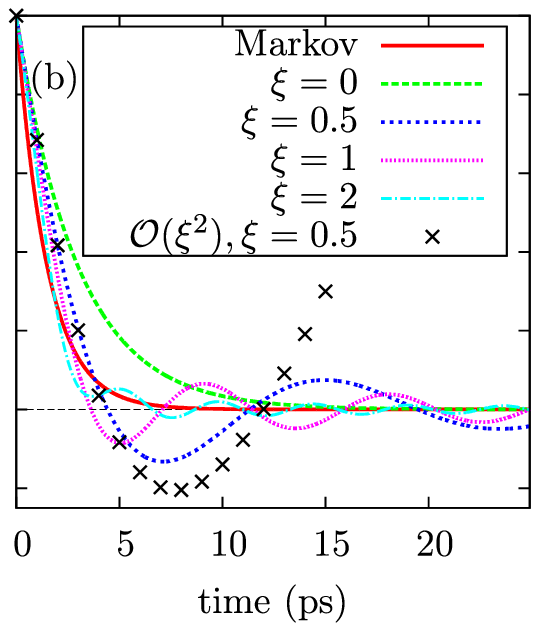}\;\;\includegraphics{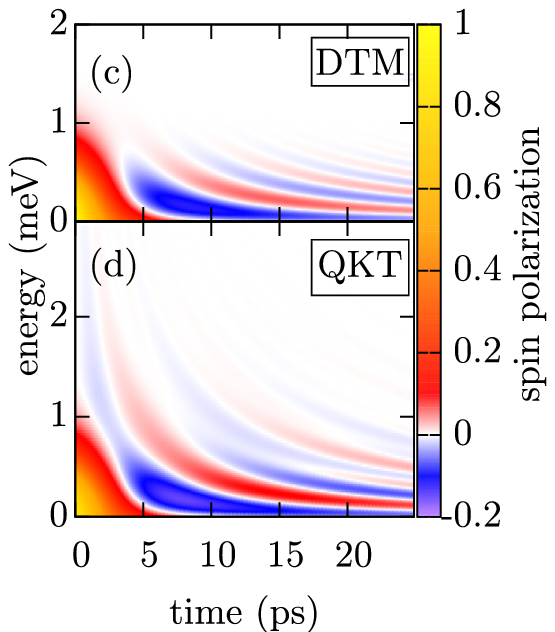}
\end{minipage}
\caption{(a): Spin dynamics in a 4 nm wide Zn$_{0.93}$Mn$_{0.07}$Se quantum 
well with $\eta\approx0.67$ ps$^{-1}$ 
with Gaussian excitaton at $\hbar\omega=0$ and
standard deviation $\Delta=0.4$ meV (same as in Ref.~\cite{Thurn13_1}) according
to the full quantum kinetic theory (QKT), 
the differential transform method 
(DTM, from Eq.~(\ref{eq:recrel})) and the Markov limit.
(b): 
Time evloution of the spin of single electrons with fixed energies
$\hbar\omega=\hbar\eta\xi$, compared with the Markov limit and the
expression in Eq.~(\ref{eq:snOrd}) for the low-$\xi$ approximation.
The spectrally resolved time evolution of the spin polarization is
shown in (c) for the DTM calculation, and in (d) for the QKT 
(cf. Ref.~\cite{Thurn13_1}).
}
\label{fig:gauss}
\end{figure}

To check the validity of the approximation of neglecting the last term of
Eq.~(\ref{eq:anf}), we compare the DTM calculation with the results of 
a full quantum kinetic treatment. Fig.~\ref{fig:gauss}(a) shows that the
non-Markovian dynamics of the total spin 
given in Fig.~1(b) of Ref.\cite{Thurn13_1} 
can be reproduced almost perfectly with the DTM calculation. 
Also, the time evolution of an individual spin of an electron
with energy $\hbar\omega_1$ is very similar in both calculations except
for a high-energy tail appearing in the full quantum
kinetic result, as can
be seen from the spectrally resolved time evolution presented in 
Figs.~\ref{fig:gauss}(c) and (d) for DTM and QKT calculations, respectively.
This finding confirms that Eq.~(\ref{eq:idgl3}) indeed captures the main non-Markovian
features of the full quantum kinetic theory.

Fig.~\ref{fig:gauss}(b) shows the results of the DTM calculation for different
values of $\xi$. For $\xi=0$, the dynamics is given by an exponential 
decay with half the rate $\eta$, as can be seen also in the
low-$\xi$ approximation in Eq.~(\ref{eq:snOrd}). For larger values of $\xi$, 
the decay rate approaches $\eta$ and oscillations start to
appear whose amplitudes eventually decrease for even larger values 
of $\xi$, where the time evolution converges to the exponential decay of the
Markov limit $(\omega_1\to\infty)$. Thus, the non-Markovian features
are only present if the approximation $\xi\gg\eta$ breaks down, i.~e., 
if the excited electrons are spectrally close to the band edge, where
the characteristic energy scale is given by $\hbar\eta$.

This can easily be understood if another derivation of the Markov limit
starting from Eq.~(\ref{eq:anf}) is considered. If the assumption of 
a vanishing memory is made and on the r.~h.~s. the functions 
$s_{\omega}(t')$ are evaluated
at $t'=t$, we can first integrate over
$dt'$ and then over $d\omega$. Calculating the first interal gives
\begin{eqnarray}
\ddt s_{\omega_1}(t)=-\frac{\eta}{\pi}\int\limits_0^{\omega_{BZ}}d\omega 
\frac{\sin[(\omega_1-\omega)t]}{\omega_1-\omega}
\bigg[s_{\omega_1}(t)+\frac 14 \big(s_\omega(t)-s_{\omega_1}(t)\big) \bigg].
\label{eq:anf_markov}
\end{eqnarray}
Using again the fact that 
$\lim_{t\to\infty}\frac{\sin[\Delta\omega t]}{\Delta\omega} \to \pi
\delta(\Delta\omega)$, one again ends up with the Markov limit.
For finite time $t$, however, the integral kernel is not yet contracted to 
a $\delta$-distribution and the finite integral limits cut off tails of the
$\frac{\sin[\Delta\omega t]}{\Delta\omega}$ function. This cut-off is 
particularly significant, if the peak of the integral kernel, which is given
by $\omega_1$ is close to one of the integral limits.

Furthermore, it can be seen in Fig.~\ref{fig:gauss}(b) that 
the low-$\xi$ approximation in Eq.~(\ref{eq:snOrd}) yields reasonable 
results for $\xi=0.5$ for the initial exponential decay while it 
fails to reproduce the long-term oscillations.

\section{Conclusion}
The non-Markovian overshoots and oscillations in the time evolution of the
carrier spins in DMS found in a quantum kinetic theory can be reproduced 
by integro-differential equation of a much simpler form that also simplifies
the interpretation considerably. A differential transform method (DTM) is 
employed to solve the resulting equation and allows to find closed-form 
expressions for low excitation energies of electrons.

It is found that a non-exponential behaviour of the time evolution of the
electron spin is only present for electrons excited close to the band edge,
where the decay predicted by the rate and the oscillations
with frequency corresponding to the electron energies 
take place on the same time scale.
Technically, this is due to the fact that a sinc-function that converges to a 
$\delta$-distribution in the Markov limit is cut off by the band edge.
It is noteworthy that similar time evolutions have also been found in different 
setups, e.~g., for the hole spin dynamics  due to 
phonon scattering in a GaAs quantum well when the scattering rate is close to
the phonon frequency\cite{ZhangWu07}. 

\ack
We acknowledge the support by the Deutsche Forschungsgemeinschaft
through the Grant No. AX 17/9-1.

\end{document}